# Constructing Metallic Nanoroad on MoS$_2$ Monolayer via Hydrogenation

Yongqing Cai,[a] Zhaoqiang Bai[b], Hui Pan[c], Yuan Ping Feng[b], Boris I. Yakobson*[d] and Yong-Wei Zhang*[a]

[a] Institute of High Performance Computing, 1 Fusionopolis Way, Singapore, 138632. E-mail: zhangyw@ihpc.a-star.edu.sg
[b] Department of Physics, National University of Singapore, 2 Science Drive 3, Singapore 117542.
[c] Faculty of Science and Technology, University of Macau, Macau SAR, China
[d] Department of Mechanical Engineering and Materials Science, Department of Chemistry, and the Smalley Institute for Nanoscale Science and Technology, Rice University, Houston, Texas 77005, USA. E-mail:biy@rice.edu

Monolayer transition metal dichalcogenides recently emerge as a new family of two-dimensional material potentially suitable for numerous applications in electronic and optoelectronic devices due to the presence of finite band gap. Many proposed applications require efficient transport of charge carriers within these semiconducting monolayers. However, how to construct a stable conducting nanoroad on these atomically thin semiconductors is still a challenge. Here we demonstrate that hydrogenation on the surface of MoS$_2$ monolayer induces a semiconductor-metal transition, and strip-patterned hydrogenation is able to generate a conducting nanoroad. The band-gap closing arises from the formation of in-gap hybridized states mainly consisting of Mo 4$d$ orbitals, as well as the electron donation from hydrogen to the lattice host. Ballistic conductance calculations reveal that such a nanoroad on the MoS$_2$ surface exhibits an integer conductance, indicating small carrier scattering, and thus is ideal for serving as a conducting channel or interconnect without compromising the mechanical and structural integrity of the monolayer.

## Introduction

Spatially confined electrical conduction at the nanoscale has been explored extensively for a number of applications in electronic or optoelectronic devices.[1] An effective approach to achieving this functionality is to construct a heterostructure by artificially breaking the periodicity of the host materials. By doing so, conducting carriers can be confined within the interface or sandwiched layer.[2,3] Through chemical engineering by atomic or molecular decoration like H passivation, confined carrier transport can also be achieved on the surfaces of otherwise intrinsically semiconducting or insulating materials. In fact, surface metallization was achieved in H-passivated SiC[4] by chemical attack and steric interaction, and in H-passivated SrTiO$_3$,[5] diamond[6] and ZnO[7,8] by electron doping from chemisorbed hydrogen species. Atomic line conduction in H-passivated Si surface was also demonstrated recently by selective removal of hydrogen atoms to create dangling bond rows.[9,10]

Monolayer MoS$_2$, a member of monolayer transition metal dichalcogenides (TMDs), consisting of an Mo atomic array sandwiched between two S atomic arrays in a trigonal prismatic arrangement, has attracted a great deal of attention due to its intriguing properties complementary to graphene.[11] The recent surge of interest in MoS$_2$ stems from its potential usage in flexible electronics and optoelectronic devices due to the presence of a finite band gap of about 1.8 eV, in contrast to the zero gap of graphene.[12] The change from indirect to direct bandgap due to the dimension reduction gives rise to strong exciton effect evidenced by strong photoluminescence emission.[13] Moreover, the lack of inversion symmetry in the monolayer along with strong spin–orbit coupling (SOC) enables efficient control of valley and spin, and allows the fabrication of novel valleytronic devices.[14,15] For practical applications in electronics and photonics, however, it is important to know how to construct stable conducting interconnects on MoS$_2$ monolayer for confined carrier transport, how various system factors affect the mobility and lifetime of carriers, and ultimately how to control the parity, content and transport of carriers.

MoS$_2$ has been well-known for its high hydrogen adsorption capacity and its ability to activate hydrogen molecule for hydroprocessing catalysis.[16,17] The surface-sorbed atomic hydrogen is believed to move from the edge sites to the basal plane through a replenishing mechanism.[18,19] In this study, we use first-principles calculations within density functional theory (DFT) to investigate the energetics of hydrogenation on the MoS$_2$ monolayer and the resulting change in electronic properties with aim to explore the possibility of the surface metallization and the construction of conducting channels. Clearly the motivation to hydrogenate MoS$_2$ here is in stark contrast to that to hydrogenate graphene: hydrogenating graphene causes its band gap opening—a metal to insulator transition—which can be used to create insulating ribbons or clusters on graphene.[20,21] Our ballistic conductance calculation shows that hydrogenation-induced metallization is viable on MoS$_2$ monolayer and strip-patterned hydrogenation can generate a nanoscale conducting nanoroad. The current approach for surface metallization on monolayer MoS$_2$ may be extended to other TMD systems.

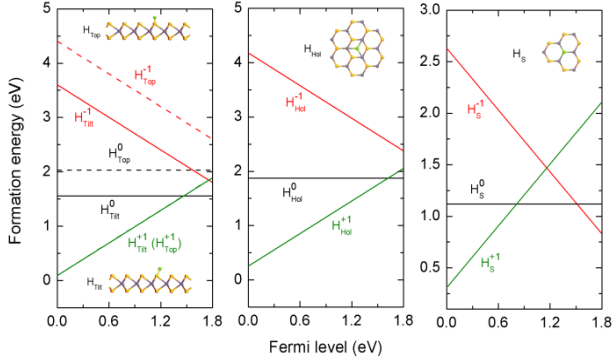

Figure 1. Defect formation energy as a function of the Fermi energy level. The insets show the various adsorption structures for H species formed on MoS$_2$ monolayer. The yellow, gray, and green balls represent S, Mo and H atoms, respectively.

## Computational details

Our DFT calculations are carried out using the plane wave code Vienna ab initio simulation package (VASP)[22] within the framework of DFT with the generalized gradient approximation (GGA). Spin-polarized calculations using the projector augmented wave method with the Perdew-Burke-Ernzerhof functional (PAW-PBE)[23] and a cutoff energy of 400 eV are adopted. For single and double H atoms adsorption, we use a $5 \times 5$ supercell based on the hexagonal primitive cell. The convergence in enegetics is tested by using a $6 \times 6$ supercell. The strip-patterned surface is created by forming H chains on MoS2 basal plane where the separation of imaging H chains in the neighboring cell is larger than 20 Å. The k point samplings are $1 \times 1 \times 8$ and $1 \times 1 \times 6$ for nanoroads along zigzag and armchair directions, respectively. Structures are relaxed until the Hellmann-Feynman force on each atom becomes less than 0.01 eV/Å. To evaluate the effect of the delocalized defective states on the H mobility of diffusion on the surface, both GGA calculation and hybrid functionals HSE06[24] in which part of the semilocal exchange-correlation functional is substituted by the Hartree-Fock exchange, are performed with the nudged elastic band (NEB) approach.

## Results and discussion

### Single H hydrogenation

Before discussing the massive hydrogenation, we first investigate the energetics and stability of MoS$_2$ hydrogenated by single H atom. Here, we consider several adsorption possibilities of such hydrogenation (Figure 1). The H atom can be located either above the basal plane directly atop an S atom (H$_{Top}$) or in a tilted structure (H$_{Tilt}$) where the S-H bond deviates 47.9 degree from the surface normal and aligns along the armchair direction. The H atom can also be adsorbed at the center in the hollow position defined by the equilateral triangle formed by three contiguous Mo atoms (H$_{Hol}$). In addition, it is possible that an H atom substitutes an S atom (H$_S$), which is equivalent to a defect complex consisting of an S vacancy and an interstitial H atom (V$_S$ + H$_i$). We have also tested the binding of H$_2$ molecule, which is found to have a very weak physisorption. Hydrogen atom in multiply bonded states with surface S atoms in the form of HS$_x$ ($x$>1) structure is found to be unstable and thus will not be studied here.

To address the stability of these hydrogenated configurations with differently charged states, we calculate the defect formation energy as a function of the Fermi level. The formation energy ($E^f$) for each configuration in the charge state of $q$ is defined as

$$E^f = E(\text{H:MoS}_2) - E(\text{MoS}_2) - \sum_i n_i \mu_i + q(E_\text{F} + \Delta V) \quad (1)$$

where $E(\text{H:MoS}_2)$ and $E(\text{MoS}_2)$ are the total energy of H adsorbed system ( H:MoS$_2$ ) and the equivalent supercell containing only monolayer MoS$_2$, respectively. The defect is formed by either adding or removing $n_i$ atoms for the $i$ type of atom with a chemical potential of $\mu_i$. $E_\text{F}$ is the Fermi level of MoS$_2$ monolayer without H adsorption. $\Delta V$ is a correction term to align the reference potential between the H:MoS$_2$ and MoS$_2$ systems. Figure 1 shows the formation energy under S-poor condition under which $\mu_\text{s}$ is bounded by the formation of Mo in the monolayer.

For hydrogenation above the anionic plane, the stable geometry depends on the charge state of the adsorbed H species. For neutral H dopant, the H$_{Tilt}$ structure is more stable and its formation energy is about 0.5 eV lower than that of the H$_{Top}$ structure. The calculated H-S bond length for H$_{Tilt}$ and H$_{Top}$ structures is 1.42 and 1.41 Å, respectively. However, for +1 charged H atom with a protonic character, the initial configuration with a tilted H reverts back to the atop structure after structural relaxation. Therefore, H$_{Tilt}$ (+1) and H$_{Top}$ (+1) share the same structure and the corresponding $E_F - E^f$ curves coincide with each other. The bond length between hydrogen atom and neighboring S atom is 1.34 Å. The transition level of (+1/0) of H$_{Tilt}$ is found to be located at around 0.3 eV below the conduction band minimum. Compared to the H$_{Tilt}$ structure, the H$_{Hol}$ defect has a higher formation energy for both the neutral and +1 states. For all the configurations, the negatively -1 charged H species are predicted to be unstable and the H atom clearly acts as an effective donor which contributes to n-type conduction of the MoS$_2$ host.

In Figure 1c, we calculate the formation energy $E^f$ of the H$_S$ defect where the single H atom occupies the center position of the V$_S$. The H atom is bonded to three Mo atoms with the bond length of around 2.0 Å. In contrast to the above-mentioned H adsorbates with a strong preference for proton or neutral H dopant, the H$_S$ defect affords a possibility of forming a negatively charged state under n-type doping condition and the formation energies of $H_S$ (0) and $H_S$ (-1) defects are clearly smaller than that of the tilted case. The reason is that the interstitial H atom forms strong bonds with the three neighboring Mo atoms. The negatively charged state of $H_S$ dopant can be stabilized by the $V_S$ defect through transferring the excess electron of the H atom to partially reduce the neighboring Mo atoms. While $H_S$ complex is important for $V_S$-rich surface, we do not consider it in the following section as we assume H decorating on a nearly perfect surface. The concentration of S vacancy on this type of surface is normally low due to a nearly equilibrium growth of $MoS_2$ layer.[25]

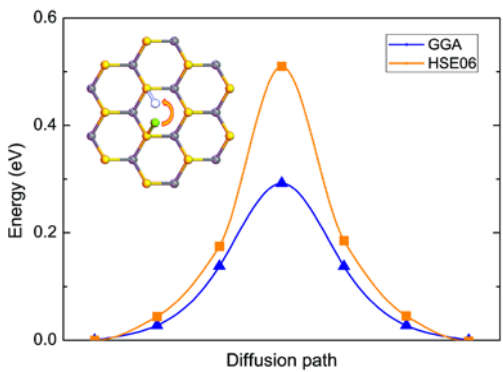

Figure 2. Activation energy of hydrogen atom diffusing on $MoS_2$ monolayer. NEB diffusion barrier is calculated for tilted H hopping between two neighboring S atomic sites. In the inset, the yellow, gray, and green balls represent S, Mo and H atoms, respectively.

In the following, we calculate the vibrational frequency for the adsorbed H species, which enables the experimental identification of the H species on the surface, where the data about charged H is still lacking.[26] According to the above result, both $H_{Tilt}$ (0) and $H_{Top}$ (+1) defects are possible due to a smaller $E^f$, depending on the Fermi level of the system. The H-S stretching frequency, obtained through solving the Hessian matrix obtained by finite difference in energy for several displacive structural configurations, is 2017 and 2487 cm$^{-1}$ for $H_{Tilt}$ (0) and $H_{Top}$ (+1), respectively. These results are consistent with the observation of vibrational bands at 2000 and 2500 cm$^{-1}$.[19,27] The blue shift of the H-S stretching frequency for +1 charged H is due to the removal of one electron from the defective level in the band gap, which is found to be an antibonding state.

Figure 2 shows the diffusion barrier by NEB calculation for an $H_{Tilt}$ defect moving to an end state with H atom bonded to the neighboring S atom. The diffusion barrier is found to be 0.29 eV using GGA method and 0.51 eV using hybrid functional HSE06. Considering the different treatments in the degree of the delocalization of the electronic states between these two approaches, the real activation barrier should be in between 0.29 and 0.51 eV. This value is much smaller than that of hydrogen diffusing on graphene, which is about 1.25 eV (GGA value, see the Supporting Information). The relatively low diffusion energy barrier, which facilitates $H_2$ spillover, may be the underlying reason for $MoS_2$ to act as an efficient hydrogen acceptor and active catalysis for numerous chemical processes.[16]

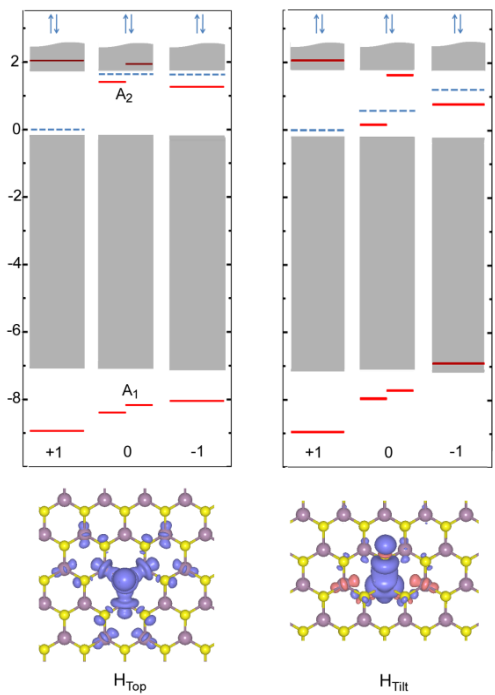

Figure 3. Energy diagram for the charged H species ($H_{Top}$ and $H_{Tilt}$ with -1, 0, and +1 states) on monolayer $MoS_2$. The blue dashed lines are the Fermi level, and the shadowed areas correspond to the conduction and valence bands of monolayer $MoS_2$. The bottom panels present the isosurface of the spin density around the H centers for the neutral H species.

We next discuss the defective levels arising from the adsorbed H species. Figure 3 shows the energy diagram of the hydrogen-derived states for $H_{Top}$ and $H_{Tilt}$ defects carrying -1, 0, and +1 charges. For both defects, two defective levels labeled as $A_1$ and $A_2$ states are identified. To understand the detailed characters, we calculated partial density of states (PDOS) and identified the percentage contribution of each atomic orbital to each level. The $A_1$ state, which is located below the valence band, is a bonding state, consisting of 21%-1$s$ of H, 52%-3$p$ of S, and 27%-4$d$ of Mo, whereas the in-gap $A_2$ state is an antibonding state, consisting of 5%-1$s$ of H, 21%-3$p$ of S, and 74%-4$d$ of Mo. These results suggest a strong hybridization of Mo states in the conduction band. For the neutral defect, the $A_2$ level, which accommodates one electron, is singly occupied and slightly spin-split. For the +1 and -1 charged cases, adding or removing one electron from the neutral system shifts these levels. As reflected by the plots of isosurface of the spin density (Figure 3), the $A_1$ state for $H_{Top}$ is found to distribute over its three neighboring Mo atoms and is more extending than that for $H_{Tilt}$, where the distribution is mainly localized at one Mo atom. Due to a larger on-site Coulomb repulsion, the more localized character in $H_{Tilt}$ $A_1$ state leads to a larger shift in the levels when the dopant is negatively charged.

For applications in electronic and optoelectronic devices, H desorption should be suppressed under electronic or photonic excitations.[28,29] The energy difference between the A$_1$ and A$_2$ is 9.5 eV and 7.8 eV for the singly occupied neutral H$_{Top}$ and H$_{Tilt}$, respectively, and increases to 11 eV for +1 charged defect. Both values are comparable to the •-•* electronic transition of 8 eV for Si-H bond in hydrogenated Si surface.[30] The relatively large separation of the bonding state and antibonding state suggests that the strong stability of H binding is able to prevent H desorption under photonic or electronic excitation for normal electronics applications.

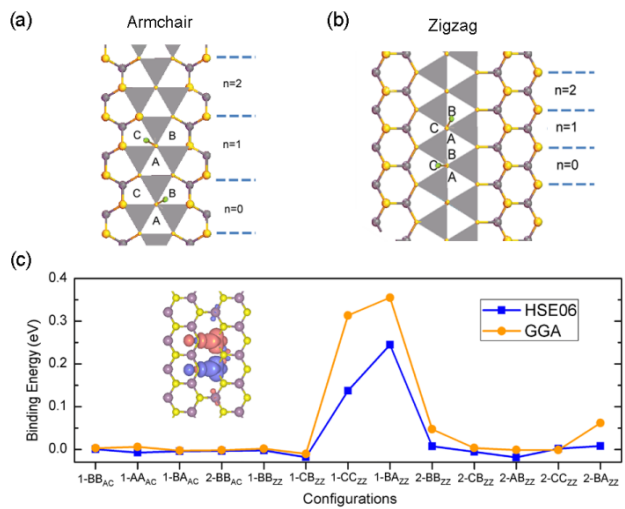

Figure 4. Coupling between two H atoms on the MoS$_2$ monolayer. Possible adsorption configurations with two H atoms aligned along armchair and zigzag directions in the hexagonal lattice are shown in (a) and (b), respectively. Each H atom can occupy one of the three vacancies (labeled as A, B, and C) around its bonded S atom. Shadow triangles representing the prisms occupied by Mo are free of H adsorption. (c) Binding energy between two adsorbed H atoms for various configurations calculated by both GGA and HSE06 methods. Inset shows the isosurface for the spin density for the 1-CC$_{ZZ}$ configuration with antiferromagnetic state.

**Dihydrogenated MoS$_2$**

It is interesting to investigate whether there is a coupling between two adsorbed H atoms, and if there is, how such coupling depends on the distance and the relative orientation of the two H atoms. To estimate this effect, we consider homolytic adsorption of an H dimer on the basal plane where each of the H adatoms forms an H-S bond. In the present work, we neglect the heterolytic adsorption, which forms one -HS group and a metal hydride, based on our above calculation, due to the estimated larger formation energy for the adsorption of the hollow H adsorption. Experimental results also indicate that there is no evidence of Mo-H bonds and only H-S bonds are present on the surface.[31] Various adsorption configurations are possible concerning the different tilted directions of H-S bonds and distances between the two atoms. Figure 4a and b show the configurations of adsorbed H atoms aligned in armchair (AC) and zigzag (ZZ) directions, respectively, according to the arrangement of their two respectively bonded S atoms. For each case, one H atom is always located in the original cell ($n$=0), and the other is placed in the $n$th (n>0) unit cell as labeled in Figure 4, and both H atoms can be tilted in the A, B, or C voids around their respectively bonded S atoms. The binding energy between two H atoms adsorbed along AC branch with one H atom located in the original cell and tilted in A direction, and the other located in the $n$th cell and tilted in B direction, for example, is denoted as $E(n$-BA$_{AC})$ and calculated using the following formula:

$$E(n\text{-BA}_{AC})= 2E(1H)-E(0H)- E(2H) \qquad (2)$$

where $E(m$H$)$ ($m$=0,1,2) denotes the total energy of the system containing $m$ H atoms.

The calculated binding energies for the various configurations are compiled in Figure 4c. It can be seen that the coupling along AC direction is very small for all the configurations. In contrast, for ZZ-direction adsorption, there exists a strong interaction with a significant dependence on the distance between the two H atoms. The two configurations with strongest coupling, that is, 1-CC$_{ZZ}$ and 1-BA$_{ZZ}$ with binding energies of 0.31 and 0.36 eV (GGA value), respectively, are identified. The 1-CC$_{ZZ}$ structure adopts a configuration with the two parallel H-S bonds normal to the ZZ direction and is denoted as "tilted-tilted" configuration. This structure is found to have an antiferromagnetic coupling, which is about 0.19 eV more stable than the ferromagnetic state. An magnetic moment is found on each Mo atom (±0.68 μB) near the end of a reversely extended line formed by the H-S bond. The opposite alignment of the magnetic moment associated with this "tilted-tilted" H dimmer may be the underlying mechanism for the recently reported ferrimagnetism in proton irradiation in MoS$_2$.[32] For 1-BA$_{ZZ}$ structure, it has a slightly larger binding energy which results from the electronic stabilization and structural relaxation. This structure, which initially has two tilted H atoms, is relaxed to a configuration with one H atom having an H$_{Top}$ configuration and the other having an H$_{Tilt}$ configuration, and is therefore denoted as a "tilted-top" configuration. According to our above result, the positive charged H$_{Tilt}$ (+1) state adopts an H$_{Top}$ configuration. This structural transition is due to an energy gain with a partial charge transfer from the defective level (see Figure 3) of one neutral H$_{Tilt}$ atom to the lattice host and the other H atom, thus leading to a quenched magnetic moment and the "tilted-top" configuration. On the other hand, the charge transfer process is accompanied by a strong structure distortion, which costs energy. A balance is reached according to the tradeoff of these two factors. We also compare the results using the GGA and HSE06 methods. The hybrid HSE06 method, which includes larger part of the degree of the localization of the electrons than GGA, predicts an overall smaller coupling value due to a smaller overlapping of these more localized electronic states.

**Strip-patterned hydrogenation**

Next, we explore the formation of conducting nanoroads on MoS$_2$ (H:MoS$_2$) via strip-patterned hydrogenation. The nomenclature of the H:MoS$_2$ nanoroads follows that of the graphene nanoribbons.[33] After the increasing uptake of H, the isolated defective levels (Figure 3) associated with a single H adsorption evolve to delocalized and metallic bands. The enhanced occupation of the 4d Mo conduction band due to the donated excess electrons from H atoms causes a strong lattice distortion arising from the Jahn-Teller effect. The "tilted-top" configuration of adjacent dual hydrogenation observed above is not formed in the H:MoS$_2$ nanoroad and all the hydrogen atoms form the H-S bond tilted from the surface normal. This is because the cost of lattice deformation energy in the "tilted-top" configuration outweighs the decrease in the on-site energy due to the electron transfer between neighboring atoms.

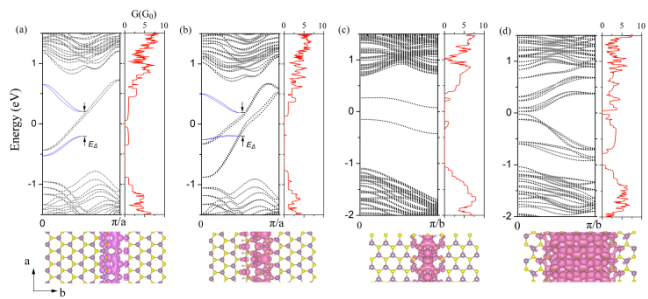

Figure 5. Band structures and ballistic conductance of H:MoS$_2$ nanoroads. Upper left panels present the band structures relative to the Fermi level for H:MoS$_2$ nanoroads along ZZ direction for (a) and (b), and AC direction for (c) and (d). In (a) and (b), the H:MoS$_2$ hybrids have 1 and 2 H chains, respectively. The blue solid lines correspond to H-derived bands for a (×2) supercell showing an induced band gap ($E_\Delta$) due to Peierls instability with doubling the size of the supercell in the chains direction. In (c) and (d), the H:MoS$_2$ nanoroads contain 1 and 8 H chains, respectively. Upper right panels present the ballistic conductance spectrum. The bottom panels show the isosurface of electron density through integrating the occupied defective states between 0 and -1 eV. Note that both the quantum conductance and isosurface plots are calculated using a double unit cell for ZZ H:MoS$_2$ nanoroad and a unit cell for AC H:MoS$_2$ nanoroad.

Figure 5 shows the band structures by taking into account SOC for H:MoS$_2$ nanoroads along ZZ and AC directions with different widths. For the ZZ H:MoS$_2$ nanoroads containing one H chain and two H chains (Figure 5 a and b), the band structures based on primitive cell have one and two bands crossing the Fermi level, respectively, suggesting a metallic conduction. For one-dimensional (1D) metallic behavior, it is necessary to examine the stability of the metallic system and the possibility of gap opening due to symmetry breaking upon slight external perturbations. Indeed, the metallic behavior in the 1D zigzag H:MoS$_2$ chain subjects to Peierls instability, and doubling the size of the supercell in the ZZ direction of H chains induces a gap of about 0.3 eV. For the AC H:MoS$_2$ nanoroads (Figure 5 c and d), in contrast, the band structure of the primitive cell shows a small gap. All the electronic bands simply follow a direct band-folding behavior and the band gap does not change when a larger supercell is adopted. For the AC nanoroad containing one H chain (Figure 5 c), two bands are identified in the band gap, which can be assigned as the in-gap defective levels corresponding respectively to the spin up and down electrons in the case of isolated H$_{Tilt}$ dopant as shown in Figure 3. Our constraint collinear spin-polarized calculation (not shown) indeed reveals that the spin-down band is empty whereas the spin-up band is singly occupied. As more hydrogenated chains are created on the surface (Figure 5d), more defective levels are formed, and thus the band gap is decreased further. It is expected that metallization can be realized in the H:MoS$_2$ nanoroad with further increasing the hydrogenation strip width.

The upper right panels of Figure 5 (a)-(d) show the calculated ballistic conductance ($G$) for the corresponding AC and ZZ nanoroads. The ballistic conductance is calculated by ATK code[34] within the nonequilibrium Green's function (NEGF) formalism. A double-• polarized basis and a cutoff energy of 200 Ry for the grid integration are adopted in transport calculations. The quantum transmission is calculated in the zero-bias regime by including self-energies for the coupling of the semi-infinite H:MoS$_2$ lead to the scattering region. It can be seen that the H rows formed on the MoS$_2$ surface can serve as conducting channels and realize coherent electronic conduction. These nanoroads allow for a perfect transmission of electrons, which is demonstrated by the integer ballistic conductance, indicating small electronic scattering. In consistence with the electronic gap in the band structures, the transmission spectra also have a gap around the Fermi level. Due to the different methods (plane-wave and atomic-orbital basis) used, the transmission gaps can be slightly different with band gaps. Upon application of a moderately biased voltage, integration of the transmission spectrum within the bias window will lead to a finite conducting current along the strip.[35]

It should be noted that all of the defective bands, except for the case of the AC nanoroad containing one H chain, show a strong SOC effect with a significant splitting between the time-reversal coupled states. Considering a relatively small SOC for electrons occupying 1s-H and 3p-S orbitals, this splitting must arise from a hybridization of 4d manifolds of the Mo atoms. Indeed, in Figure 5, the occupation of the cation atomic plane is reflected in the isosurface plot of the integrated defective states in the band gap. As the defective bands, which result in the transmission of the electrons, contain a large component of the Mo 4d empty orbitals, the conducting behavior of these H channels is unlikely to be affected by defects like disordering induced by the different H orientations or point defects like hydrogen vacancy formed on the nanoroad. The metallicity in the hydrogenated MoS$_2$ monolayer revealed here resembles the conduction on the hydrogenated Si surface where the conducting channels arise from dangling bonds.[9,10] A disadvantage of the dangling-bond-creation approach is that isolated dangling bonds tend to pair up, thus reducing the reliability and stability of surface metallization.[36,37] In the current approach, hydrogen atoms act as electron donors to the Mo atoms and the conducting channels consist of the hybridized states of mainly 4d Mo and 3p S.

As in the case of graphane, where precise control of H passivation on the graphene surface is highly important for applications, the efficiency and robustness to reach controllable hydrogenation on MoS$_2$ surface determines its functionality and performance. While the amount of H adsorption on MoS$_2$ can be expressed as $H_xMoS_2$ ($x<1$),[18] the accurate value of $x$ on the surface is still unknown. Although we have shown that a small amount of defects like H vacancies will not degrade the metallic behavior of the hydrogenated $MoS_2$ strip (see the Supporting Information), the ability to control H coverage is still a critical issue in any device applications. Previous studies have shown that Ni, Pt and Co-promoted $MoS_2$[19,38,39] has a much larger uptake of H compared with unpromoted $MoS_2$. Another possible approach to ensure an enough uptake of H is through $p$-type doping of as-deposited $MoS_2$ surface before hydrogenation to take advantage of the compensation effect of H species as the donor center. The enhancement of H content by this doping strategy is well-known for B-doped H:Si surface[40] and H passivation on other III-V semiconductors.[41] As the present approach for metallicity of H:$MoS_2$ system relies on the charge transfer between H atoms and sandwiched cation layer, it is expected this conducting behavior may also exist in other TMD $AX_2$ systems, in which the electronegativity of cation in the middle layer dominates the ability of the charge transfer. Indeed, considerable amount of hydrogen was found to be able to adsorb on $WS_2$,[42] and metallic conduction was observed in $H_xNbSe_2$ ($0<x<0.53$).[43] More recently, a H-incorporated $TiS_2$ ultrathin nanosheet with a ultrahigh conductivity is reported where a tunable conductivity depending on the H content is demonstrated.[44]

## Conclusions

In summary, we have investigated the energetics, electronic structures and surface metallization of the hydrogenation on monolayer $MoS_2$. We demonstrate that the hydrogenation creates defective levels in the band gap of $MoS_2$ and these defective states, which are mainly composed of Mo 4d orbitals, form a continuous band upon strip-hydrogenation on the surface, and create a one-dimensional conducting nanoroad, enabling the construction of interconnects in $MoS_2$-based nanoelectronic devices. It should be noted that this strategy is different from the metallization of H:Si[9,10] or H:SiC[4] surface, where quasi-conducting channels are related to the unsaturated dangling bonds through removing H from the surface. Here, we show that the ballistic conduction of carriers can be achieved by tunneling through H(1$s$)-S(3$p$)-Mo(4$d$) hybridized states, which are mainly localized in the middle Mo layer. An important feature of the H:$MoS_2$ structure as conducting interconnects in nanocircuit is that the van der Waals gap between $MoS_2$ layers and between $MoS_2$ monolayer and substrate can suppress possible current decay arising from the current leakage from the top layer to the interior bulk or substrate. Therefore, the present work provides a viable route to construct conducting nanoroads in $MoS_2$-based nanocircuits.

## Acknowledgements

This work was supported by the A*STAR Computational Resource Centre through the use of its high performance computing facilities.